\documentclass[twocolumn,superscriptaddress,prl,showpacs,preprintnumbers,amsmath,amssymb]{revtex4-1}
\usepackage{amsmath}
\usepackage{amssymb}
\usepackage{textcomp}
\usepackage{dcolumn}
\usepackage{graphicx}
\usepackage{bm}

\begin{document}
\title{Quantitative prediction of the phase diagram of DNA-functionalized nano-colloids}

\author{Bianca M. Mladek} \email{bianca.mladek@univie.ac.at} 
 \affiliation{Department of
  Chemistry, University of Cambridge, Lensfield Road, Cambridge, CB2
  1EW, UK}

\affiliation{Max F. Perutz Laboratories
  GmbH, University of Vienna, Department
  of Structural and Computational Biology, 1030 Vienna, Austria}

\author{Julia Fornleitner} \affiliation{Institute for Complex Systems,
  Forschungszentrum J\"ulich, 52428 J\"ulich, Germany}

\author{Francisco J. Martinez-Veracoechea} \affiliation{Department of
  Chemistry, University of Cambridge, Lensfield Road, Cambridge, CB2
  1EW, UK}

\author{Alexandre Dawid} \affiliation{Universit\'{e} Joseph Fourier
  Grenoble 1/CNRS,  Laboratoire Interdisciplinaire de Physique UMR 5588, Grenoble, 38041, France}

\author{Daan Frenkel} \email{df246@cam.ac.uk} \affiliation{Department of
  Chemistry, University of Cambridge, Lensfield Road, Cambridge, CB2
  1EW, UK}


\date{\today}

\begin{abstract} 

  We present a coarse-grained model of DNA-functionalized colloids
  that is computationally tractable. Importantly, the model parameters are solely based on
  experimental data. Using this highly simplified model, we can
  predict the phase behavior of DNA-functionalized nano-colloids
  without assuming pairwise additivity of the inter-colloidal 
  interactions.  Our simulations show that for nano-colloids, the
  assumption of pairwise additivity leads to substantial errors in the
  estimate of the free energy of the crystal phase.  We compare our
  results with available experimental data and find that the
  simulations predict the correct structure of the solid phase and
  yield a very good estimate of the melting temperature. Current
  experimental estimates for the contour length and persistence length
  of single-stranded DNA sequences are subject to relatively large
  uncertainties. Using the best available estimates, we obtain
  predictions for the crystal lattice constants that are off by a few
  percent: this indicates that more accurate experimental data on
  ssDNA are needed to exploit the full power of our coarse-grained
  approach.

\end{abstract}


 \maketitle 

Nature provides spectacular examples of complex, functional systems
that self-assemble from small, pre-fabricated units. The grand
challenge in nano-material design is to imitate this phenomenon to
construct complex, functional materials.  To achieve this goal we must
be able to design suitable nano-sized building blocks and to create
conditions that cause these entities to self-assemble into the desired
target structures. DNA-based building blocks---which take advantage of
the selective binding of bases on complementary strands to guide
assembly---offer attractive model systems to explore self-assembly
strategies (see e.g.~Refs.~\cite{Bas01,Ald08}).  For instance,
colloidal particles can be functionalized with a short single-stranded
(ss) DNA sequence tethered to an inert polymeric
``spacer''~\cite{Mir96}. These ``sticky ends'' on two different
colloids may then either bind directly to each other via complementary
sequences or via a ssDNA linker sequence introduced in solution,
allowing the colloids to form three-dimensional structures. The
properties of these building blocks, temperature, pH and ionic
strength of the parent solution determine if, and on what time scale,
self-assembly takes place.\\ In view of the vastness of this ``design
space''~\cite{Gee10}, a careful selection of experimental conditions
is crucial: experiments on
nano-~\cite{Par08,Nyk08,Mac09,Xio09,Mac10,Cig10} and micron-sized
DNA-functionalized colloids (DNACs)~\cite{Bia05,Kim06} show that
self-assembly of spatially ordered structures requires considerable
fine-tuning: under most experimental conditions, amorphous aggregates
form, even if the ordered (crystal) phase is thermodynamically
stable. The experiments show that the stability of crystals depends
crucially on temperature ($T$) and on the length and flexibility of
the spacers.  Of the crystal structures experimentally observed, the
bcc and fcc lattices are the most stable, while to date little
progress has been made with self-assembly of DNAC crystals that are
more complex than cubic.\\ Therefore, it is desirable to have
theoretical and numerical guidance in selecting optimal conditions for
self-assembly. However, an ``ab-initio'' approach to study the phase
behavior of DNACs {\it quantitatively}, using many-particle
simulations, goes well beyond the state-of-the-art.  Existing
theoretical studies are based on the assumption that the interaction between DNACs is pair-wise additive~\cite{Tka02}, 
and numerical simulations of DNACs typically employ highly simplified, ad-hoc coarse-grained models that also assume
pairwise  additivity  of interactions~(see
e.g.~Refs.~\cite{Luk04,Pie05,Sta06,Boz08,Dre09,Mar10,Leu11,Sca11,Kno11}).
An exception is the work of Rogers {\it et al.}~\cite{Rog11,Mog12}.
However, the approach followed in their paper, whilst suited to
compute pair potentials between micro-sized DNACs, would become
computationally prohibitively expensive for nano-sized DNACs where, as
we will show below, it is not warranted to assume pairwise additivity
of the inter-colloidal interactions. What is more, those systems
that have been shown most promising for crystallization, are
nano-DNACs covered with DNA strands in an intermediate length regime,
where the DNA strands are too long to be approximated within the
rod-picture \cite{Leu11}, but too short to make use of polymer scaling
laws. As a consequence, existing models cannot be used to predict the
stability of crystal phases of nano-DNACs.\\ In this Letter, we
present a quantitative numerical approach to predict the thermodynamic
stability and phase behavior of nano-DNACs based on a coarse-graining
procedure free of fitting parameters.\\ We validate our
coarse-graining approach on a system that has been studied extensively
in experiments~\cite{Nyk08}, namely a symmetric, binary mixture of
gold nano-colloids of a radius $R_C \sim 6$nm which were grafted with
$\sim$ 60 ssDNA strands. Colloidal species A and B only differ in the
sequence of their sticky ends, which mediate the binding between
colloids A and B. In the experiments of Ref.~\cite{Nyk08}, it was
observed that systems with ssDNA strands of more than 50 nucleotides
crystallize into CsCl structures. In our study, we therefore focus on
their system of DNACs functionalized with ssDNA strands of 65
nucleotides, 15 of which were responsible for binding. These DNACs
were experimentally found to crystallize for temperatures below
$T_m^{\rm exp}$=62.5$^{\circ}$C.\\ In order to arrive at a
computationally tractable model, we carry out a staged coarse-graining
procedure. At the most microscopic (yet not fully atomistic) level, we
represent the Au colloid as hard sphere of radius $R_C$ and we model
the ssDNA strands as freely jointed, charged chains with a Kuhn length
of 1.5nm~\cite{Zha01,Tin97}, and using an inter-base distance for
ssDNA of 0.43nm \cite{Tin97}.  We stress, however, that the values reported
in these (and other) papers are subject to considerable error bars,
and are likely to be sensitive to both the physical
conditions of the solution and
the precise DNA sequence.  Hence, we should expect that these
inaccuracies will translate into errors in computed characteristic
length scales, such as the lattice constants of DNAC crystals. As we
show below, this is precisely what we observe.\\ The charge carried by
the ssDNA's backbone is assumed to be distributed equally among the
vertices of the freely jointed chain.  The vertices interact through a
Debye-H{\"u}ckel potential that depends on $T$, and on the dielectric
constant and ionic strength of the solvent~\cite{Zha01}.  We assume
that the grafted DNA is distributed uniformly on the Au surface and
does not diffuse~\cite{Str94}.\\ In spite of the simplicity of this
model, it has a large number of degrees of freedom per colloid ($\sim$
3000). As a consequence, simulations are only feasible for relatively
small systems.  In order to be able to treat DNA hybridization in
systems containing many colloids, we therefore develop the next level
of coarse-graining: a ``core-blob'' model in which each sticky end is
coarse-grained to a polymer ``blob'', while the bare colloid and the
remaining segments of all strands constitute an effective colloidal
``core''.  In this way, we reduce the degrees of freedom per colloid
tenfold. We determine the parameters and interaction potentials that
characterize the core-blob model in Monte Carlo simulations of the
underlying microscopic model~\cite{Mla10}. Further, we allow for DNA
hybridization in the core-blob model. The binding probability of
complementary DNA strands is computed on the basis of the tabulated
hybridization free energy of two complementary ssDNAs in
solution~\cite{Mar05}, which is directly based on experimental data.
Details of the coarse-graining procedure and the Monte Carlo scheme
used to carry hybridization moves will be presented elsewhere
\cite{MlaXX}.\\
\begin{figure}
\includegraphics[width=3.3in]{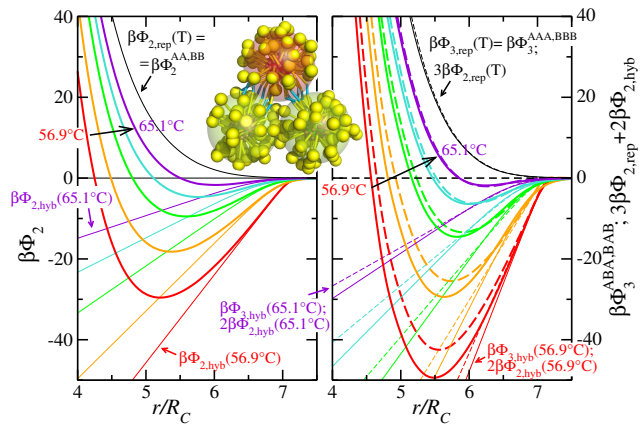}
\caption{(Color online) Left: The pair interaction $\Phi_2^{\rm AB}(r, T)$
  (bold lines) is the sum of the $T$-independent steric repulsion
  $\Phi_{2, \rm rep}(r)$ (black thin line) and the strongly
  $T$-dependent attractive hybridization energy $\Phi_{\rm 2,
    hyb}(r,T)$ (colored thin lines). Right: The three-body effective
  interaction $\beta \Phi_3^{\rm ABA}(r)$ (bold solid lines) of two A and one B
  colloids arranged in an equilateral triangle (see inset) compared
  with the sum of the two-body contributions $3\beta \Phi_{2,{\rm
      rep}}+2 \beta \Phi_{2,{\rm hyb}}$ (bold dashed lines). Further,
  the repulsive and the attractive hybridization contributions to the
  interactions are shown explicitly in thin lines (two-body: dashed
  lines, three-body: solid lines). Both: Results are shown at
  56.9$^{\circ}$C, 59.1$^{\circ}$C, 61.4$^{\circ}$C, 63.2$^{\circ}$C,
  and 65.1$^{\circ}$C. Inset: A simulation snapshot of 1 B colloid
  (red solid sphere) interacting with 2 A (green solid spheres) at
  59.1$^{\circ}$C and $r=5.75R_C$. Free blobs are drawn as yellow
  spheres, while hybridized sticky ends are shown as blue rods. The
  translucent spheres indicate the extension of the core.}
\label{fig:lvl2_3}
\end{figure}
In what follows, we will use the core-blob model to predict the phase
behavior of DNACs.  However, we first go one step further in
coarse-graining and compute the effective pair potentials
$\Phi_2^{\aleph}(r,T)$ between two DNACs ($\aleph = \{$AA,BB,AB$\}$)
as a function of the distance $r$ and of $T$. $\Phi_2^{{\rm
    AA,BB}}(r,T)$ accounts for the steric repulsion $\Phi_{2,{\rm
    rep}}$ between the colloids, which can be determined as detailed
in \cite{Mla10}. In the AB case, there is an additional attractive
hybridisation interaction $\Phi_{2,{\rm hyb}}$, which is proportional
to the amount of hybridized DNA strands and is calculated by
thermodynamic integration as described in Ref. \cite{Leu11}. We find
that $\Phi_{2,{\rm rep}}$ is nearly constant between 25 and
75$^{\circ}$C, while $\Phi_{2,{\rm hyb}}$ is strongly $T$-dependent
(Fig.~\ref{fig:lvl2_3}).  For $T \lesssim T_m^{\rm exp}$,
$\Phi_2^{{\rm AB}}$ therefore develops a strongly $T$-dependent
minimum at distances $r \sim 5.25$ to 6$R_C$. For the system presented
here, a $T$-difference of only 5.2$^{\circ}$C (62.1 to
56.9$^{\circ}$C) leads to a difference in minimum in $\Phi_2^{\rm AB}$
of $\sim 20k_BT \sim 13$kcal/mol.  The difficulty to make high-quality
crystals of DNACs is related to this strong temperature dependence of
$\Phi_2^{\rm AB}$: defects in growing crystals can only anneal if $T$
is just below the melting temperature.  At lower $T$, bonds are very
stable and dense aggregates, once formed, cannot equilibrate.\\ We are
unaware of direct measurement of the pair potentials of
DNA-functionalized {\it nano}-colloids, hence we cannot validate our
model at this level.  However, numerically, we can test whether the
assumption of pairwise additivity of $\Phi_2^\aleph$, as typically
employed in theoretical studies, is warranted.  To this end, we
compute the interaction $\Phi_3^\aleph(r,T)$ ($\aleph = \{{\rm
  AAA,BBB,ABA,BAB}\}$) between a triplet of colloids arranged on the
vertices of an equilateral triangle with sidelength $r$
(inset~Fig.~\ref{fig:lvl2_3}). We find that $\Phi_3^{\rm AAA,BBB} =
\Phi_{3,{\rm rep}}$ is well represented by $3 \Phi_{2,{\rm rep}}$ for
all $T$ considered. For the ABA (=BAB) case, additivity would imply:
$\Phi_{3}^{ABA}(r,T) = 3 \Phi_{2,{\rm rep}}(r,T) + 2 \Phi_{2,{\rm
    hyb}} (r,T)$. As can be seen in the right panel
of~Fig.~\ref{fig:lvl2_3}, even for temperatures close to $T_m^{\rm
  exp}$ this relation is only fulfilled at large separations $r\gtrsim
6R_C$.  The reason why additivity fails in this case is that different
$B$ colloids compete for the same DNA on $A$.  This effect is not
accounted for in $\Phi_2$, which measures the average amount of bonds
in an {\it isolated} AB pair and hence overestimates the attraction
between AB pairs in a trimer~\cite{footnote}.\\ Turning our attention
to multi-particle systems, we study the thermodynamic stability of
various crystalline phases. Using the pair potentials, we first
identify candidate equilibrium crystal structures via a genetic
algorithm similar to the one used in Refs.~\cite{Got05,For08}. This
approach, which neglects positional entropy, suggests that the CsCl
structure is most stable close to the melting density, while a NaTl
structure is predicted to be stable at higher densities.
Experimentally, the distinction between these two structures is not
straightforward as X-ray scattering only probes the arrangement of the
Au cores which is the same for CsCl and NaTl crystals. We also
considered CuAu, NaCl, `straight' hcp (s-hcp), ZnS, AuCd and
substitutionally disordered CsCl and CuAu structures.
\begin{figure}
\includegraphics[width=3.3in]{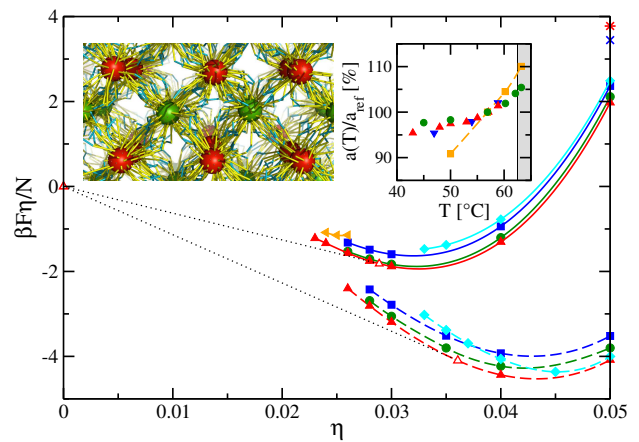}
\caption{(Color online) The dimensionless free energy $\beta \eta F/N$
  at $T=56.9^{\circ}$C as function of the colloidal volume fraction
  $\eta$ shows that the CsCl structure is the most stable one. Dashed
  lines: pair potential approach; solid lines: core-blob approach.
  CsCl:$\blacktriangle$, s-hcp:$\bullet$, CuAu:$\blacksquare$,
  NaTl:$\blacklozenge$, disordered CsCl:$\ast$, disordered
  CuAu:$\times$, liquid:$\blacktriangleleft$. The common tangents
  between the equilibrium CsCl crystal and the dilute vapor are shown
  as dotted lines for both models. Right inset: the lattice constant
  $a$ as function of $T$ in $^{\circ}$C as measured in the experiments
  (heating:$\blacktriangle$, cooling:$\blacktriangledown$; data from
  Ref.~\cite{Nyk08}) and as obtained from simulations with the
  core-blob model ($\bullet$) and using the pair potential approach
  ($\blacksquare$). The data of $a$ have been scaled to the values of
  $a_{\rm ref}=a(56.9^{\circ}{\mathrm C})$ as measured with the
  respective approaches: $a_{\rm ref}^{\rm exp} =
  45.3$nm, $a_{\rm ref}^{\rm core-blob}=39.7$nm and
  $a_{\rm ref}^{\Phi_2}=36.9$nm. In the shaded
  region crystals are not stable in experiments. Left inset: a cut
  through a simulated CsCl structure at $\eta=0.024$ and
  $T=56.9^{\circ}$C. For better visualization, only hybridized links
  are drawn.}
\label{fig:free_energy}
\end{figure}
For all crystal structures and also the fluid phase, we compute the
free energy $F$ for $T\lesssim T_m^{\rm exp}$ via thermodynamic
integration within the core-blob model (Fig.~\ref{fig:free_energy}).
The CsCl phase is found to have the lowest free energy for low
colloidal volume fractions $\eta$. Only slightly higher in $F$, we
find metastable s-hcp, CuAu and NaTl structures.  While the CsCl is
already mechanically stable for $\eta = 0.023$, the latter structures
are only mechanically stable for slightly higher $\eta \gtrsim
0.026-0.035$.  At high $\eta \sim 0.065$ a metastable AuCd phase
appears. The NaCl and ZnS structures are mechanically unstable for all
$\eta$ considered.  For $\eta \gtrsim 0.07$, the NaTl structure
(composed of two interpenetrating diamond structures) competes---as
predicted---with the CsCl structure; however, this is in a density
regime where the validity of our model is not guaranteed.  Moreover,
such dense crystals cannot easily be prepared in experiments.\\ We
stress that the formation of the low-density crystals is due to the
DNA links between colloids. At much higher $\eta$ the crystal
structure will be dictated by excluded volume interactions rather than
DNA links. Then, close-packed structures such as CuAu and s-hcp should
be more stable than the more open structures that dominate at lower
$\eta$.\\ We also consider substitutional disorder and find that
strongly disordered crystal structures are only mechanically stable
for $\eta \gtrsim 0.05$ and less stable than the corresponding ordered
structures. However, {\it some} substitutional disorder is inevitable:
at 56.9$^{\circ}$C, the free-energy cost of a {\it single} AB exchange
in an otherwise ordered crystal structure ranges from $\Delta F =
0.5k_BT$ ($\eta = 0.024$) to 1.6$k_BT$ ($\eta = 0.05$) for CsCl and
from $\Delta F = 0.1k_BT$ ($\eta = 0.026$) to 1.0$k_BT$ ($\eta =
0.05$) for CuAu. We note that substitutionally disordered crystals
have been observed both in experiment~\cite{Par08} and in
simulations~\cite{Kno11}.  The present results suggest that slightly
disordered structures are metastable but kinetically arrested for $T
\lesssim T_m^{\rm exp}$ while strongly disordered crystals will melt
at low $\eta$.\\ In Fig.~\ref{fig:free_energy}, we further compare $F$
as obtained from the core-blob approach with the corresponding
quantity obtained employing the pair potentials. The assumption of
pairwise additivity of $\Phi_2$ leads to a serious under-estimate of $F$. Nevertheless,
for $T \lesssim T_m^{\rm exp}$ this approach provides a
reasonable estimate of the range of mechanical stability of the
various crystal structures and predicts the same phase order as the
core-blob model, albeit in a more compressed $\eta$ range.\\ We also
used the core-blob model to test whether the colloidal suspension
could undergo a transition between a dilute and a concentrated
unordered phase below the freezing density and for $T \lesssim
T_m^{\rm exp}$.  We find that this is not the case: the CsCl-structure
directly coexists with a very dilute suspension~\cite{Mar11}. Using
the common-tangent construction (see~Fig.~\ref{fig:free_energy}), we
can determine the volume fraction---and thereby the lattice constant
$a$---of the CsCl structure at coexistence.  Comparison of the
experimentally determined $a$ at various $T$ with the values obtained
from the core-blob approach (right inset~Fig.~\ref{fig:free_energy})
shows that the latter approach can predict the correct temperature
dependence of the lattice spacing. This is not the case for the pair
potential approach, which predicts an excessive contraction of the
crystals. Both computational approaches predict lower values of $a$
than observed experimentally. The core-blob approach underestimates
$a$ by $\sim$12\%, a discrepancy to be expected in view of the
incomplete experimental information on the contour length and
persistence length of ssDNA.  With better experimental data, we expect
that the core-blob model will yield a quantitative prediction of the
DNAC crystal lattice constants. The predictions of the lattice
constants as obtained by use of pair potentials are qualitatively
wrong - in particular this approach incorrectly predicts the
temperature dependence of the lattice spacing.  We can further compare
the experimentally determined crystal melting point with the melting
point of simulated crystal slabs in equilibrium with the dilute
vapor. In these simulations, we find that crystals melt for $T \gtrsim
64.3(5)^{\circ}$C, which matches well with the experimentally
determined melting temperature $T_m^{\rm exp} = 62.5(3)^{\circ}$C for
the 65 nucleotide system of Ref.~\cite{Nyk08}. The good quantitative
agreement between simulation and experiment in melting temperature is
significant as the core-blob model contains no {\it a posteriori}
adjustable parameters.\\ The present approach has been developed to
model DNACs with direct hybridization between sticky ends. However,
our approach can be generalized to describe binding via linker
sequences and it can be adapted to describe asymmetric and
polydisperse systems, to name but a few examples.  Although some
adjustment of the underlying microscopic model of DNACs is needed to
mend the accuracy of length predictions, our approach offers a path to
the computer-aided design of DNA-functionalized building blocks that
could be used to construct truly complex self-assembling structures.

\begin{acknowledgments}
We thank B.~Capone, S.~Angioletti-Uberti,
B.~M.~Mognetti, W.~Jacobs and G.~Day for useful
discussions. BMM acknowledges EU funding (FP7-PEOPLE-IEF-2008
No.~236663) and the MFPL VIPS program (funded by the BMWF and City of
Vienna). AD was supported by an EMBO longterm fellowship. DF and FJMV
acknowledge support of ERC Advanced Grant 227758. DF acknowledges a Wolfson
Merit Award of the Royal Society of London and EPSRC Programme Grant EP/I001352/1.
\end{acknowledgments}


\bibliographystyle{apsrev}

\end{document}